\documentclass{IEEEtran}
\usepackage{cite}
\usepackage{amsmath,amssymb,amsfonts}
\usepackage{algorithmic}
\usepackage{array}
\usepackage[caption=false,font=normalsize,labelfont=sf,textfont=sf]{subfig}
\usepackage{url}
\usepackage[hidelinks]{hyperref}
\usepackage{xurl}
\usepackage{graphicx,color}
\usepackage{float} 
\usepackage{textcomp}
\usepackage{tabularx}
\usepackage{booktabs}
\begin{document}

\title{
A Hardware-in-the-Loop Experimental Testbed using Air Conditioners for Grid Balancing}

\author{Oluwagbemileke E. Oyefeso, \textit{Student Member}, \textit{IEEE}, Drew A. Geller, Ioannis M. Granitsas, \textit{Member}, \textit{IEEE}, Duncan S. Callaway, \textit{Senior Member}, \textit{IEEE}, and Johanna L. Mathieu, \textit{Senior Member}, \textit{IEEE}
\thanks{This work was supported by ARPA-E Award DE-AR0001061.}} 

\maketitle

\begin{abstract}
Driven by the need to offset the variability of renewable generation on the grid, development of load control is a highly active field of research.  However, practical use of residential loads for grid balancing remains rare, in part due to the cost of communicating with large numbers of small loads and also the limited experimentation done so far to demonstrate reliable operation.  To establish a basis for the safe and reliable use of fleets of compressor loads as distributed energy resources, we constructed an experimental testbed in a laboratory, so that load coordination schemes could be tested at extreme conditions.  This experimental testbed was used to tune a simulation testbed to which it was then linked, thereby augmenting the effective size of the fleet. Modeling of the system was done both to demonstrate the experimental testbed's behavior and also to understand how to tune the behavior of each load.  Implementing this testbed has enabled rapid turnaround of experiments on various load control algorithms, and year-round testing without the constraints and limitations arising in seasonal field tests with real houses. Experimental results show the practical feasibility of an ensemble of small loads contributing to grid balancing.
\end{abstract}

\begin{IEEEkeywords}
Air Conditioners, Experimental Testbed, Frequency Regulation, Hardware-in-the-Loop, Load Control
\end{IEEEkeywords}

\section{INTRODUCTION}
New deployments of renewable energy resources have increased rapidly over the last decade, driven both by new policies to address climate change and by economics as the price of renewable generation rapidly decreases~\cite{cali}.  Given that the fastest growing forms of renewable generation -- solar and wind -- are variable and intermittent, more fast-response storage and load control capacity will be needed to maintain a reliable grid as the generation mix evolves with a reduced fraction of schedulable/controllable generation.  Load coordination, however, is difficult because different types of electrical loads have different characteristics and customer requirements, so that the reliability, consistency, and performance of aggregations of loads is uncertain.  For load aggregations to gain acceptance as grid balancing resources and avoid liability with both customers and system operators, it is necessary to test and debug control schemes to failure in an experimental environment before moving to field testing with real houses.

In this paper we consider aggregations of air conditioners (ACs) that switch on/off to maintain a temperature near a setpoint. Their power consumption can be controlled by sending switching commands or temperature settings directly to the thermostats.  One reason for selecting ACs is because of their ubiquity, with ACs present in almost 90\% of single-family homes in the U.S.~\cite{recs}.  Another reason is that residential ACs consume more energy than other appliances, so they have a larger impact on the electricity supply-demand balance if their consumption can be shifted in time~\cite{callaway_tapping_2009}.  ACs and other compressor-based loads are more challenging to control than resistive loads like electric heaters, because compressors have a ``lockout" time between when they are shut off and can be turned on again, to guard against short-cycling.  Short-cycling risks damaging the compressor or causing it to stall and draw a high current until it overheats or trips a circuit breaker.  ACs also display a variable power draw as they start up and reach equilibrium temperatures on the hot and cold heat exchangers.  Finally, the AC's compressor draws a transient inrush current (significantly larger than normal operating current) when switched on, and the impact on transformers and distribution circuits of multiple ACs switching on at exactly the same time should be considered.

We constructed an experimental testbed at Los Alamos National Laboratory to study the effects of load control on residential window ACs in nearly constant ambient conditions. The experimental testbed consists of 20 single-zone model houses of nearly identical construction, situated within a warehouse. To ensure that these models were representative of real houses, initial open-loop experiments were run and validated against data from 47 homes in Austin, TX supplied by Pecan Street, Inc.~\cite{PSI}. Because even 20 ACs represents only a small aggregation, the experimental testbed was designed to optionally interact with a simulation of hundreds to thousands of virtual houses to boost the effective size of the testbed. Several closed-loop feedback control algorithms were tested to demonstrate the aggregation's ability to provide frequency regulation. Confidence in these integrated hardware-in-the-loop (HIL) experiments was established by comparing results from the experimental houses to the virtual houses to check that the control algorithms did not favor either set. 
Additional features of the testbed include a programmable heat source, to emulate changes in occupancy of the house, and the ability to tune the natural cycling period of ACs in the house with minor physical modifications.

The experimental testbed described here is one of a very small number of laboratory experiments that have been built at this scale, e.g.,~\cite{bindner2016,supermarket,vrettos_experimental_2018a, vrettos_experimental_2018b}.  In contrast to our testbed, \cite{bindner2016} focused on refrigerators; specifically, the authors investigate the ability of refrigerators to provide frequency regulation via experiments on 25 refrigerators. Refs.~\cite{supermarket,vrettos_experimental_2018a, vrettos_experimental_2018b} focused on commercial rather than residential buildings; \cite{supermarket} explores demand response from a commercial refrigeration system; and \cite{vrettos_experimental_2018a, vrettos_experimental_2018b} used an experimental facility~\cite{flexlab} to demonstrate frequency regulation with a commercial HVAC system. 

In summary, the contributions of our paper are fourfold. First, we design, develop, and validate an experimental testbed for testing AC load control strategies. Second, we develop an extended equivalent thermal parameter (ETP) model of ACs, which captures more of the salient characteristics of these devices, e.g., the cooling time lag. Third, we use the testbed to test three load control methods over a wide range of scenarios, using both physical and simulated ACs, demonstrating the usefulness of the testbed and the ability of ACs to provide frequency regulation. Fourth, 
we highlight the opportunities and challenges associated with testing AC load control strategies through HIL experiments. These findings may be of interest to researchers and practitioners exploring methods to test/benchmark load control strategies before pushing them to the field. These findings also highlight some of the inherent challenges in using AC aggregations for grid balancing, and some techniques for overcoming those challenges.  

In Section~\ref{HILA}, we describe the HIL architecture, and the design and features of our experimental testbed. In Section~\ref{testbed-learnings}, we explore the behavior of the experimental testbed, derive a new ETP model, and discuss experimental testbed tuning. Section~\ref{TestbedValidation} describes HIL testbed validation. The results of our HIL experiments are provided in Section~\ref{HILexperiments}, and we conclude the paper in Section~\ref{conclusion-section}.  

\section{Hardware-in-the-loop Architecture} 
\label{HILA}

\begin{figure}
    \centering
    \includegraphics[width = \columnwidth]{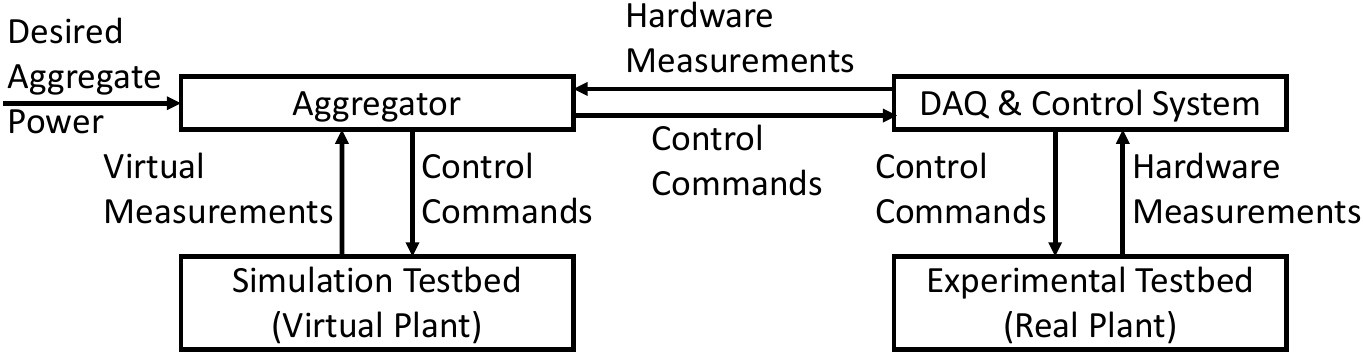}
    \vspace{-.6cm}
    \caption{Hardware-in-the-Loop experiment architecture.}
    \label{fig:ETBPA}
    \vspace{-.3cm}
\end{figure}

The HIL experiment architecture is shown in Fig.~\ref{fig:ETBPA}. An aggregator running a load control algorithm receives a desired aggregate power command from, e.g., the system operator, and sends control commands to an AC aggregation. Our testbed differentiates between the experimental testbed (20 physical model houses representing the real plant), which receive the commands via a data acquisition (DAQ) and control system, and the simulation testbed (hundreds to thousands of simulated houses representing the virtual plant), which serves to boost the number of ACs in the aggregation.  ACs within the simulation testbed are henceforth referred to as ``virtual ACs" and ACs within the experimental testbed are henceforth referred to as ``experimental ACs." Control commands and measurements are transmitted back to the aggregator via a communication system. The following subsections detail each component shown in~Fig.~\ref{fig:ETBPA}.

\subsection{Experimental Testbed Design} \label{testbed-design}
Although our model houses are not scaled from any particular house, they exhibit salient features found in typical houses, as we will show in Section~\ref{TestbedValidation}. Also, although some houses have two or more cooling zones, the testbed model houses are single-zone units.  Limiting the models to single-zone systems is not especially restrictive, as many single-family houses have single-zone central air conditioning. Also, window ACs are found widely in single rooms of houses, apartments, hotel rooms, and small offices.  

\begin{figure}
\centering
\includegraphics[width=0.45\columnwidth]{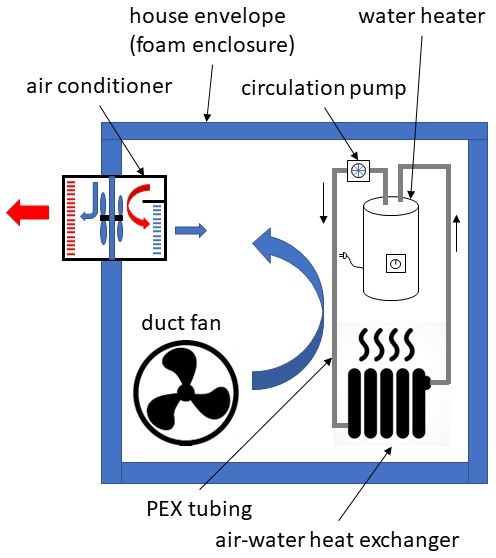} \includegraphics[width=0.49\columnwidth]{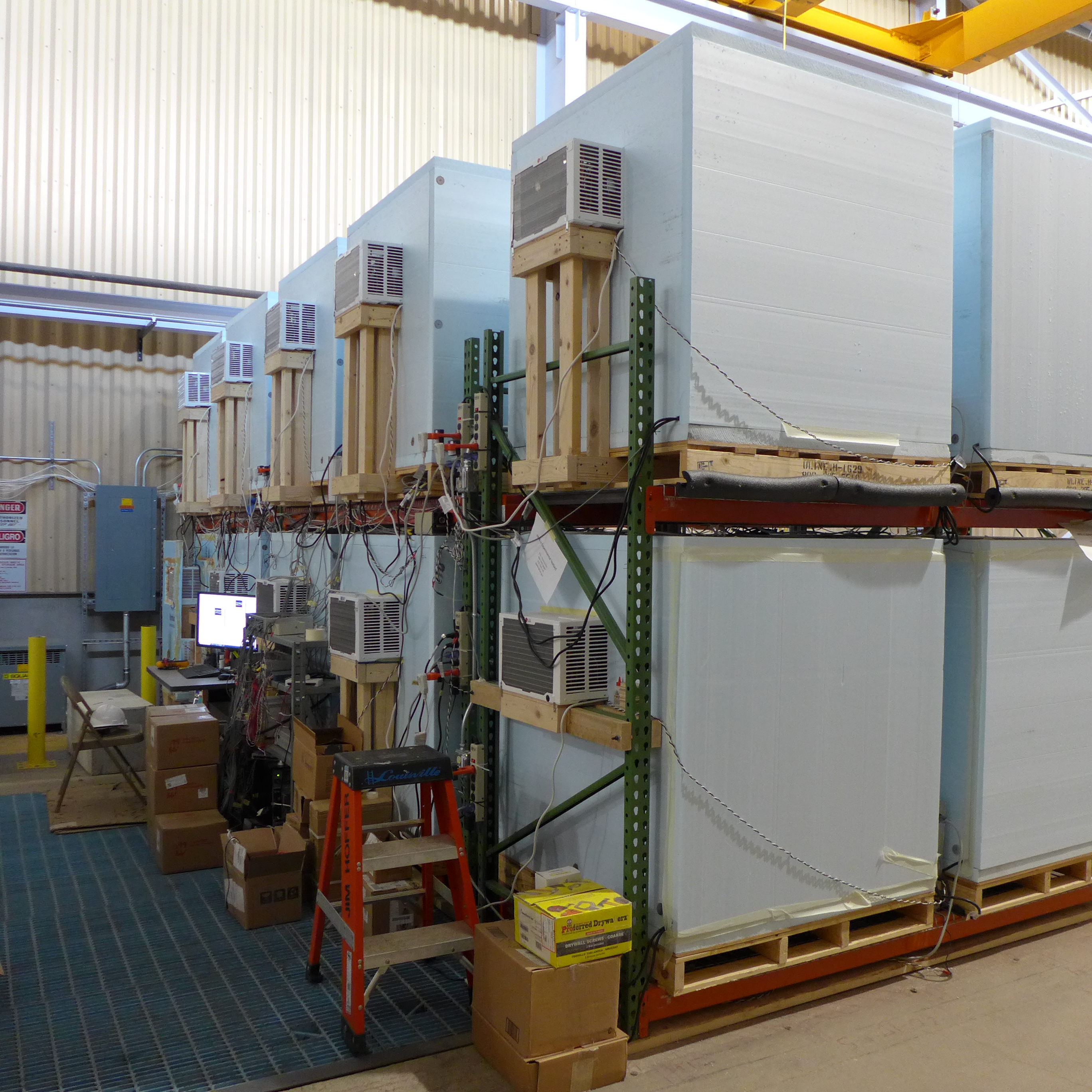}
\vspace{-.3cm}
\caption{Left: Model houses are constructed inside a four foot cubic foam box.  The internal heat source consists of a hydronic loop with an electric water heater.  A duct fan forces air through the heat exchanger and mixes the room air in addition to the AC's fan. Right: The 20 units are stacked in pallet racks, minimizing cable lengths and overall footprint.}
\vspace{-.5cm}
\label{model-house}
\end{figure}

The model houses have three subsystems: (1) the experimental AC, (2) the house envelope and environment, and (3) the thermal properties, including the heat source and circulation fan. A sketch depicting the design and a photograph of testbed are given in Fig.~\ref{model-house}. We next describe each subsystem.

\subsubsection{Air Conditioners}
The ACs used in this testbed are single-speed.  Although dual-speed and variable-speed ACs are more efficient and provide improved user comfort, they are not yet dominant in the U.S.~\cite{DOEIEDO}.  
To accommodate as many model houses as possible in the space available, we selected 5000 BTU/h window ACs, which are the smallest capacity commonly sold.  The model selected contained a mechanical thermostat, and the thermostat was easily replaced with a relay controlled externally by the DAQ computer.  Also, the ACs have a single speed fan, which was turned on all the time both to provide additional air circulation in the box and also to distinguish its power consumption from that of the compressor.  The ACs are equipped with a passive element that interrupts power if the compressor stalls and overheats, but adjustable switching delays or lockout periods were imposed in the DAQ software to prevent short-cycling the compressor.  Because the built-in mechanical thermostat was not used, the DAQ and control system provided the thermostat function in software, switching based on temperatures measured by negative temperature coefficient thermistors in each house.  

\subsubsection{House Envelope and Environment}
The insulating boundary of each house was a 1.2~m cube foam enclosure constructed from 5~cm thick extruded polystyrene foam boards with an R-value of 10.    
The humidity of the air is normally low in Los Alamos, NM with typical values around 10\% relative humidity.  Relative humidity in the lab was measured with a sensor, but the humidity was not controlled. No adjustments to the data were needed to account for humidity as no condensate was pulled from the air.

\subsubsection{Thermal Properties}
In real houses, solid materials exchange heat with the air and significantly affect the temperature dynamics through their heat capacities.  The model houses contain electric point-of-use water heaters, of either 20 or 30 gallon capacity, and this water represents the solid heat capacity normally comprised by the solid walls and furnishings.  By adjusting the heat exchange between water and air, one may effectively make the model house appear to have a higher heat capacity and thus cycle the compressor on/off less frequently, given a fixed internal heat load.  In this way, the model can behave like a real house despite its smaller size, at least in terms of its AC cycle durations.  The air and water are thermally coupled through a 0.3~m square tube-and-fin heat exchanger, and water is circulated from the water heater through cross-linked polyethylene tubing and a pump to the heat exchanger and back to the water tank.  This is similar to the hydronic loop of a residential radiant heat system.  A single-speed duct fan is placed next to the water-air heat exchanger to flow air through it and mix the air throughout the small box.  Were no heat generated in the heater, pump, and fan, the water would simply contribute to the apparent heat capacity in the house.

The water circulation pump and the duct fan are fixed electric loads enclosed in the house, and they run constantly during experiments.  They are therefore constant internal heat loads.  In contrast, the water heater is used as a programmable heat source, with the amount of heat injected adjusted by a solid-state power controller.  
For each on-off cycle, the AC must remove all heat generated in the house during that cycle, independent of the total heat capacity of the air and water.

Each water heater has its own power controller circuit, which allows for each house to have different internal heat gains.  The DAQ computer controls all the heating rates and it also allows the operator to add randomness to the heating rate (if needed to disturb phase-locking or synchronization of houses) or to make the heating rate vary over time to emulate occupant usage patterns or diurnal usage variation.

The assembly was constructed in a compact, two-level arrangement 
shown in Fig.~\ref{model-house}.  Though the construction of each model house is nearly identical, we found that the units tended to have different cooling loads and cycling behavior. This is likely due to a variety of factors, including variation in pumps, fans, and ACs; differences in water heater sizing, differences in envelope construction, differences in the location of the units in the warehouse, component positioning within the units, and differences in the location of the thermistor within the units.  The last of these has a large impact as discussed in Section~\ref{testbed-learnings}.

\subsection{Simulation Testbed Design} \label{sim-testbed}

The simulation testbed is described in detail in~\cite{Granitsas2024,code_repo}, where we also describe its integration with a field testbed in Austin, TX. Here we briefly describe its main features. The testbed, implemented in MATLAB, incorporates high-fidelity ETP AC models~\cite{GridlabD} validated against data from homes in Austin, TX~\cite{PSI}. We model AC active power consumption dependence on outdoor temperature, reactive power consumption dependence on voltage, as well as inrush current resulting in high active/reactive power draws that may occur for very short time durations after the compressor turns on. The simulation testbed also enables the analysis of system-wide distribution network effects of load control strategies by assigning each AC to a node of a taxonomy distribution feeder from~\cite{PNNL}. GridLAB-D~\cite{GridlabD} is used as a power flow solver. 
Aggregator-to-AC communication networks are modeled with configurable links, package drop rates, and delays.

\subsection{Aggregator's Load Control Algorithms} \label{controllers}

We tested three previously-developed load control algorithms. Each has the same objective of regulating the aggregate power consumption of the ACs to match a reference power command; however, each controller differs in architecture and algorithm. All controllers were implemented in MATLAB.

The first controller is a proportional-integral (PI) controller, which uses the system output (aggregate AC power consumption) to compute the reference tracking error, and subsequently the next control input according to standard PI control design~\cite{Golnaraghi2017}.  

The second control approach is the Markov model-based probabilistic control described in \cite{Mathieu2013}. The benefit of this control scheme is that it uses a model of aggregate AC dynamics to predict AC power consumption one step ahead. It uses broadcast control and can incorporate a state estimator to simplify the design and reduce the cost of the communication network. We use the version of the controller that incorporates delayed temperature dynamics from~\cite{temp-delays}.

The third control approach is the device-driven extended Packetized Energy Management (PEM) strategy described in~\cite{Leke2022}. PEM, originally introduced in~\cite{Almassalkhi2017}, allows devices to make a request to turn on for fixed-duration periods to consume a fixed amount of energy. The load aggregator coordinating the devices approves or denies the requests to track the power reference command. Ref.~\cite{Leke2022} extended this approach to accommodate compressor-based devices like ACs by including flexible turn-on periods and turn-off requests. 

\subsection{Data Acquisition and Control System} \label{CCDAQ}
The DAQ and control system serves as a communication intermediary between the aggregator's load control algorithms and the experimental testbed, as shown in Fig.~\ref{fig:ETBPA}. The system encodes and sends measurements from the experimental testbed such as house temperature and AC power consumption to the aggregator, while also processing the aggregator's control commands, both translating the command into AC-specific switching signals and checking whether or not each signal is feasible given the current on/off and lockout states of that AC. Only feasible commands are transmitted.

The DAQ and control system is implemented in National Instruments LabView. The entire array of model houses, the laboratory sensors, and the power meter monitoring of each house are managed by a single computer system.  
The LabView software includes a simple TCP server that allows it to communicate with other programs on the same computer.  Since the AC states for the experimental houses were measured in real time, the DAQ and control system clock, contained in the LabView experiment, determined the latching of time steps, such that the aggregator's controller would step along with the physical experiment. 

\subsection{Communication System} 
\label{Comm}
Communication and data exchange of control commands and virtual measurements between the aggregator and simulation testbed were implemented in MATLAB. However, the communication and data exchange between the aggregator and the experimental ACs is achieved via the DAQ and control system using TCP server port connections. 
Switching control commands from the aggregator are first encoded as digital signals and then further encoded as JavaScript Object Notation (json) strings. This is all done in MATLAB, and sent to a specific TCP port address, i.e., the receiving TCP port address of the DAQ and control system. Likewise, the aggregator receives encoded measurements from the experimental ACs via the DAQ and control system, at its own specific TCP port address. Measurements are subsequently parsed by decoders and data filters built into the aggregator's controller to detect and flag any corrupt data (e.g., infeasible state measurements) before being passed on to the aggregator's control algorithm. 

\section{Learnings from the Experimental Testbed} \label{testbed-learnings}

\subsection{Experimental Testbed Behavior} 
\label{testbed-behavior}
\begin{figure}
\centering
\includegraphics[width=0.98\columnwidth]{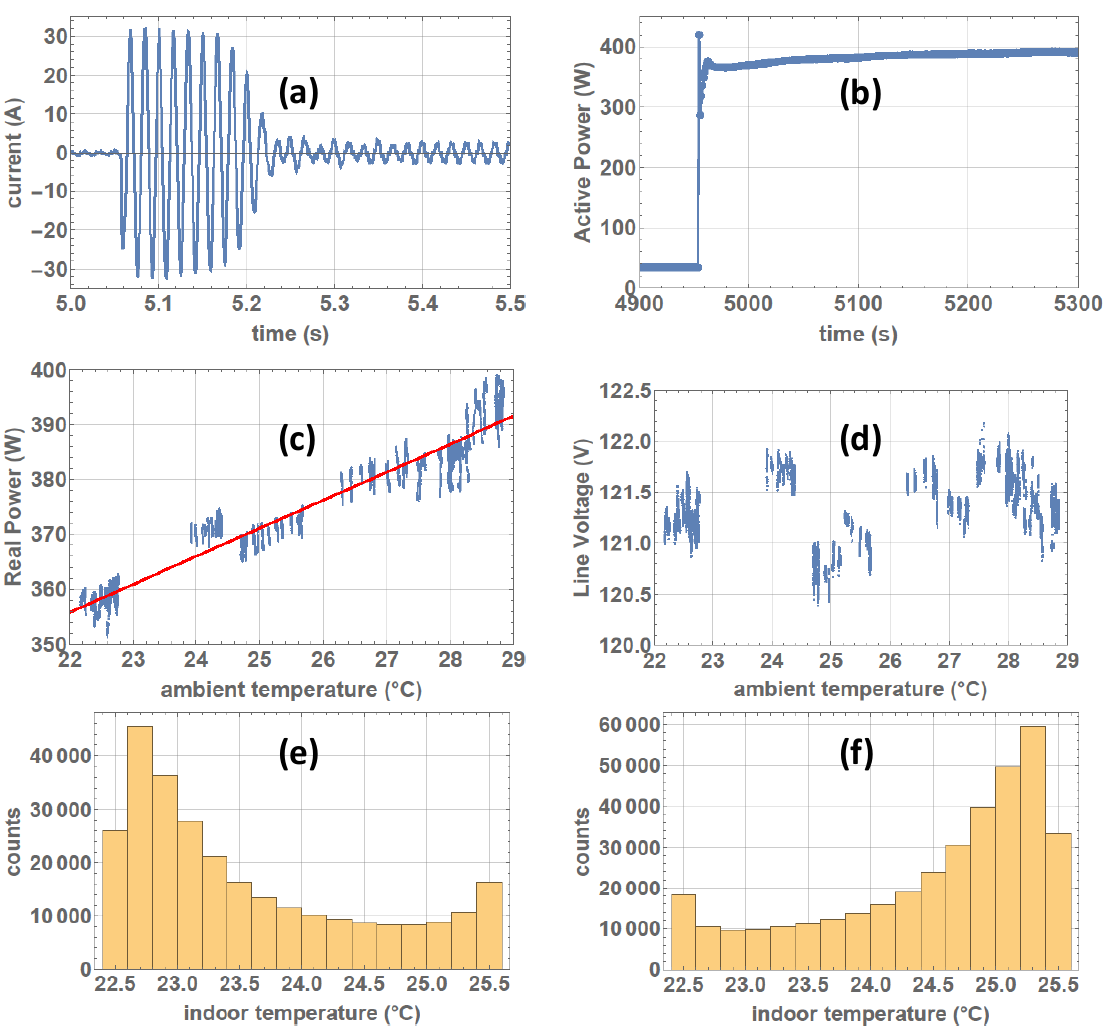}
\vspace{-.3cm}
\caption{(a) Inrush current measured for a single AC. (b) Time evolution of active power for a single AC. (c) Peak power consumption of an AC varies with ambient temperature.  (d)  The line voltage is uncorrelated with ambient temperature in the lab.  (e) Histogram of temperatures for all 20 model houses, measured every second and integrated over several hours, while ACs have their compressors on and (f) off.}
\label{fig:testbed-combined}
\vspace{-.5cm}
\end{figure}

It is useful to run experiments on physical model houses to locate phenomena that might be overlooked in simple mathematical models.  Such phenomena may cause a control algorithm that worked well in simulation to not work as expected on real systems.  For example, ACs do not simply turn on and have a flat power draw, as they are commonly modeled.  Rather, there are fast and slow features in the power consumption.  Initially, there is a spike in power due to the inrush current, because the compressor takes time to start from rest, where the stalled input impedance is small.  It generally takes 5-10 60~Hz AC cycles for this inrush current to decay, as seen in Fig.~\ref{fig:testbed-combined}(a).  The energy associated with the inrush current is a negligible contribution to the consumption of the AC; however, the inrush current may matter to the grid, because many ACs turning on simultaneously will briefly draw five or six times the steady-state current expected.  

Following the inrush transient, the power consumption briefly drops as the compressor delivers refrigerant vapor up to the expansion (or throttle) valve leading to the evaporator.  The power then rises rapidly for a few seconds and with the pressure at the throttle valve, as the vapor is pressurized and condenses to a liquid there.  Then the power begins to rise more slowly and may reach a plateau as the evaporator and condenser heat exchangers approach their minimum and maximum temperatures, respectively.  This evolution in power consumption is shown in Fig.~\ref{fig:testbed-combined}(b). The entropy generation in moving heat from low to high temperature requires work from the compressor.  Energy is also consumed by friction in the compressor and by the work done in compressing the gas into the condenser.  There are no corresponding delays in power consumption approaching zero when turning off the compressor.

The ambient temperature outside the model houses affects the house dynamics in two ways.  First, there is the heat leak through the walls of the house, which depends on the temperature gradient across the foam wall.  Higher temperatures outside the box will show up as higher duty cycles for the ACs, as they need to remove extra heat in each cycle.  In these model houses, the additional heat load was approximately 5 W/$^{\circ}$C through the walls.  Second, even if the heat leak into the box were negligible, the hot heat exchanger must now reject the same amount of heat into a higher-temperature ambient reservoir, and the compressor must work correspondingly harder.  Since the fan's flow rate is fixed and the heat transfer coefficient is unchanged, the hot heat exchanger would increase in temperature to maintain a constant temperature difference $\Delta T$ with the environment.  For these model houses, the increase in instantaneous power was $1.36\;  \rm{\%/{^\circ}C}$ as shown in 
Fig.~\ref{fig:testbed-combined}(c), which includes data from a single AC compiled for runs under identical heat injection but varying lab temperatures.  The data shown are the power when the AC is on, dropping the data immediately after the compressor starts. This increase in instantaneous power is similar, but not identical, to the change measured in~\cite{CIEE-report} and also that seen in real AC data~\cite{PSI}.

From real AC data, we usually see the line voltage drop at higher temperatures as more consumers use their ACs and increase the load on the feeder.  In the lab, though, the testbed is not large enough to affect the feeder voltage, so no voltage drop is seen. Therefore, in the lab, line voltage is uncorrelated with ambient temperature as shown in Fig.~\ref{fig:testbed-combined}(d).

For any of the houses, the instantaneous probability of finding the house at any temperature within the temperature deadband is not uniform, with the houses spending most of their time near the edges of the temperature deadband $[T_-, T_+]$.  This is shown in Figs.~\ref{fig:testbed-combined}(e) and~\ref{fig:testbed-combined}(f)
for all experimental ACs when they are on or off, respectively. This occurs because temperature evolves nonlinearly when heat is driven by temperature gradients from the environment or internal heat sources.  The mean temperature, referenced to the setpoint temperature, will vary almost inversely with the duty cycle of the AC, within the deadband limits. For example, if the injected heat is high enough that the AC is on 90\% of the time to make the thermostat reach $T_-$, then the room must be cooling very slowly as it approaches $T_-$ and the time-averaged temperature must be close to $T_-$.  If the injected heat is minimal but the room still warms when the AC is off, then the temperature approaches $T_+$ slowly and the time-averaged temperature is close to $T_+.$ 

The temperature generally shows a delayed response to turning on/off the compressor, so the temperature exhibits small excursions (as much as 0.1 $^{\circ}$C) outside the deadband when the AC changes state.  The AC itself has some thermal inertia in that the compressor must first start to cool the heat capacity of the metal evaporator before the air begins to cool; and, similarly, when the compressor halts, the condensed refrigerant continues to cool the heat exchanger briefly as the condensate vaporizes.  There is also a time constant for the mixing of the air in the room, limited by the flow rate of the circulation fans in the space ($\approx$ 12 s in these model houses).  Finally, the temperature lag will depend on the location of the room's thermometer and its proximity or attachment to nearby solid features.  These delayed temperature dynamics~\cite{temp-delays} could cause  tracking excursions in the controller, since the controller may rely on a prediction of instantaneous thermometer response to the compressor after a state change.  These features are not normally represented in ETP models~\cite{etp-model}. 

\subsection{New Equivalent Thermal Parameter Model}
\label{ETPmodeling}
We developed an ETP model to understand the main features of the model houses.  The parameters of the model were chosen to approximately represent the houses, but the model was not
calibrated using system identification to represent any particular house.  Rather, the model was used to explore some  house behaviors with respect to the air-water heat transfer and to the AC itself.  To accomplish this, the ETP model was extended by using a simple model of the AC and by introducing a temperature offset to the thermostat measurement. 

\begin{figure}
\centering
\includegraphics[width=.8\columnwidth]{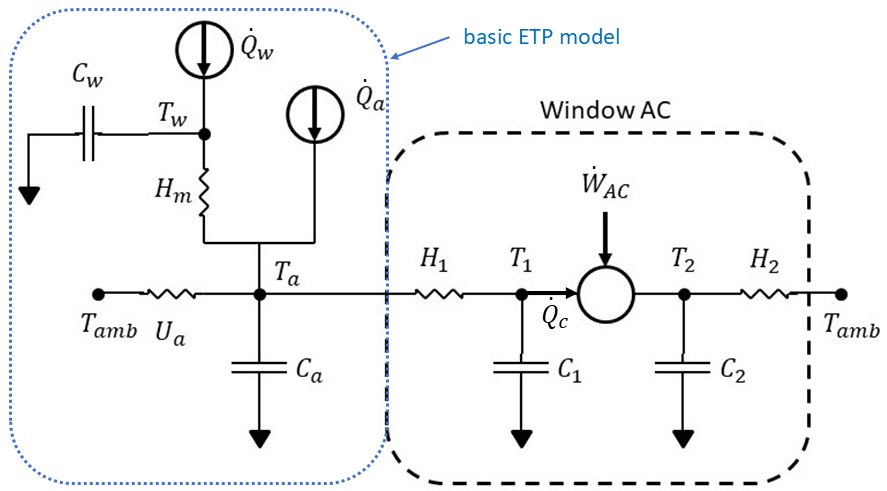}
\vspace{-.3cm}
\caption{The new ETP model of a model house.  The basic ETP model from~\cite{etp-model} is on the left. In this basic model, $\dot{Q}_a$ would represent the AC as simply a step function with heating rate $\dot{Q}_a = 0$ when the AC is off and $\dot{Q}_a = -\dot{Q}_c$ when the AC is on.  Here, though, we extended the model to explicitly include a simple model of the window AC to account for the time-varying power draw in cooling the house. Including this lossy Carnot refrigerator model introduces physics missing from the basic model, such as the time lag for the AC's heat exchangers to warm or cool and the effect of the outside temperature on the active power consumption of the AC (cf. Fig.~\ref{fig:testbed-combined}(c)).}
\label{circuit-diagram} 
\vspace{-.5cm}
\end{figure}

The new ETP model is depicted in Fig.~\ref{circuit-diagram}.  The rate of heat injection into the water is $\dot{Q}_w$.  Generally, the direct heating rate of air $\dot{Q}_a$ is negligible compared to the other heat loads since there is always a solid (e.g., windings in the duct fan) that is directly heated and transfers heat to the air from its surface.  In that case, $\dot{Q}_a \approx 0$ and the heat generation can effectively be grouped with $\dot{Q}_w$.  The outside temperature $T_{\rm amb}$ affects the system both through heat conducted through the walls, with conductance $U_a$, and also by being the temperature of the reservoir into which the AC rejects heat.

The heat flow equations for this model are
\begin{eqnarray}
    C_w \dot{T}_w & = & H_m \left( T_a -T_w \right) + \dot{Q}_w  \label {tweqn}\\
    C_a \dot{T}_a & = & U_a \left( T_{\rm amb} - T_a \right) + H_m \left( T_w - T_a \right) \nonumber \\
    & & \;\;\; +\; \dot{Q}_{\rm a} + H_1 \left( T_1 - T_a \right) \\
    C_1 \dot{T}_1 & = & H_1 \left( T_a - T_1 \right) - \dot{Q}_c \\
    C_2 \dot{T}_2 & = & H_2 \left( T_{\rm amb} -T_2 \right) + \dot{Q}_c + \dot{W}_{\rm AC},
    \label{t2eqn}
\end{eqnarray}
where $T_w$ is the water temperature, $T_a$ is the air temperature, $T_1$ is the temperature of AC's cold heat exchanger (evaporator), $T_2$ is the temperature of the AC's hot heat exchanger (condenser), $H_m, \, H_1, \, H_2$ are heat transfer coefficients, $C_w, \, C_a, \, C_1, \, C_2$ are heat capacities, $\dot{Q}_c$ is the heat removal rate of the AC, and $\dot{W}_{\rm AC}$ is the active power consumption of the AC. Both $\dot{Q}_c$ and $\dot{W}_{\rm AC}$ are defined in Appendix~\ref{appendix} and render the system of equations nonlinear. Therefore, we solve them numerically.  Specifically, we solve the system of equations until $T_a$ reaches $T_-$, then we solve them with $\dot{Q}_c =0$ until $T_a$ reaches $T_+$, and then we repeat until the piece-wise temperature trajectory reaches a steady-state cycling pattern.

\subsection{Tuning the Experimental Testbed}
\label{testbed-tuning}
\begin{figure}
    \centering
    \includegraphics[width=\columnwidth]{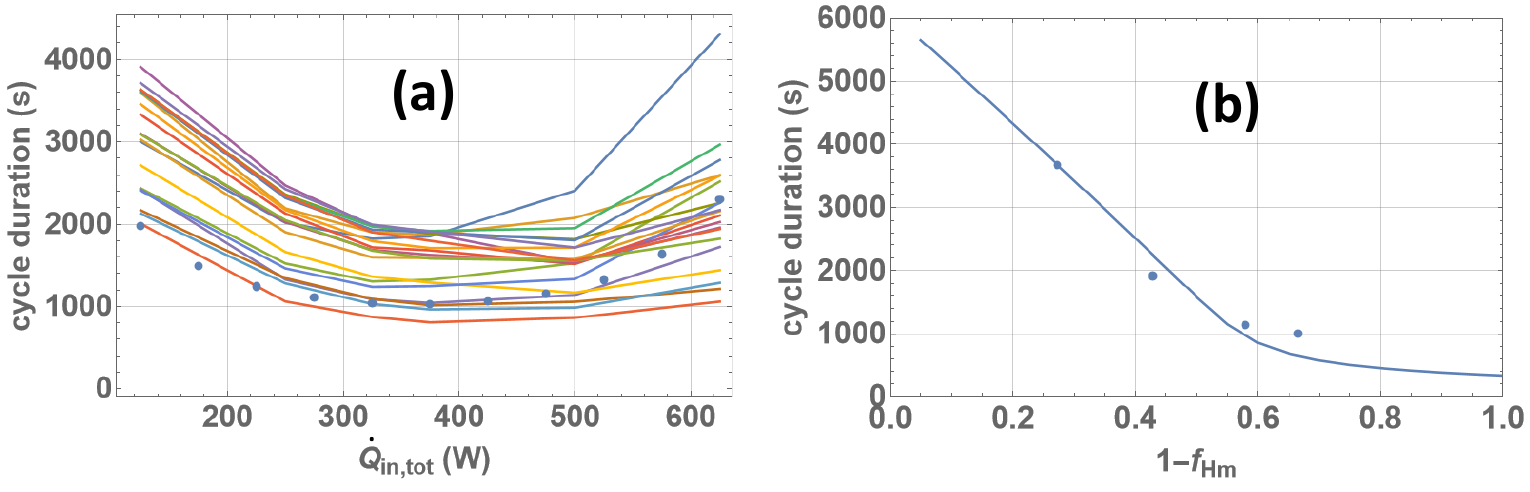}
    \vspace{-0.5cm}
    \caption{(a) Cycle duration vs.~internal heating rate for the experimental ACs (lines) and for the extended ETP model (circles).  (b) Cycle duration vs.\ thermometer placement for the extended ETP model (line).  As $1-f_{Hm}\rightarrow 1$, the thermostat uses the mixed air temperature in the middle of the model house, and the air temperature is weakly coupled to the water temperature so that the AC cycle durations are short.  As $1-f_{Hm}\rightarrow0$, the air temperature is tightly coupled to the water temperature, and the entire heat capacity of the water must heat to $T_+$ and cool to $T_-$ in each cycle.  Circles are data from an experimental AC run with its thermostat sensor at four different positions above the air-water heat exchanger.}
    \label{cycle-durations}
    \vspace{-.5cm}
\end{figure}

For validating the experimental testbed against real AC data, as will be described in Section~\ref{TestbedValidation}, we wanted the AC cycling patterns to be similar.  The real AC data varies for each house with outside air temperature, solar irradiance, occupancy, and temperature setpoints chosen by homeowners.  Using the ETP model, one can show the variation of the AC's on-time, off-time, and resulting cycle duration as a function of the heat gain in the unit.  The results from the ETP model are plotted against data from the experimental ACs in Fig.~\ref{cycle-durations}(a) and are seen to be similar in form.  For low internal heating rates, the AC cools the room quickly, but the temperature rises with time constant $\sim \left( C_a + C_w \right) \left( T_+ - T_- \right)/\dot{Q}_{\rm in,tot}$, where $\dot{Q}_{\rm in,tot}$ is the total heat injection rate from fixed and programmable sources including the fixed heating rate of the duct fan and the water pump, which together add 125~W to the programmed value.  For heating rates approaching $\dot{Q}_c$, the maximum cooling rate of the AC, the time constant for cooling diverges with $\left( C_a + C_w \right) \left( T_+ - T_- \right)/(\dot{Q}_c- \dot{Q}_{\rm in,tot})$ as $\dot{Q}_{\rm in,tot} \rightarrow \dot{Q}_c$.  The cycle duration is therefore lowest for moderate ($\approx 50\%$) duty cycles, but the range of heat injections producing low durations is quite broad.

Although the 20 houses are constructed to be nearly identical, the heating/cooling rates and resulting cycle durations were found to span a wide range.  The units with the shortest cycles were found to have higher time-averaged water temperatures.  This might suggest that one should improve the thermal contact between air and water, so that more of the water's heat capacity is effectively accessed in each thermal cycle.  One way to do this would be to increase the size of the heat exchanger, e.g., by increasing the size of the radiator in the hydronic system.  However, this requires altering the design and should not be necessary given that other identical units operate with useful cycle times, i.e., on/off times that are longer than the compressor's lockout time (3 min).

Increasing the average air flow speed through the heat exchanger is \emph{not} a solution that increases the cycle duration in a model house.  Assuming fully-developed laminar air flow in the heat exchanger, the convection coefficient $h$ is nearly constant with air speed (see (8.53) of \cite{incropera}), so that the instantaneous heat flux $\dot{q}_{HX} = h \left( T_{w} - T_{a} \right)$
would be unchanged.  In these units, the air flow is fast enough that it does {\em not} reach fully-developed thermal or velocity profiles within the depth of the heat exchanger, even though the flow is still laminar.  The ``entrance region" spans the whole thickness of the heat exchanger, and $h$ does increase with air speed.  Still, the average heat transfer coefficient is only increasing with the cube root of the velocity from its value for fully-developed laminar flow until one reaches the transition to turbulent flow~\cite{incropera}.  In spite of the increase in heat transfer rate, the thermometer used for the thermostat will see a decrease in its coupling to the water temperature.  The thermometer is immersed in the exit flow from the heat exchanger, and an increase in air speed through it results in a lower temperature difference relative to the well-mixed air in the model house because
    $\dot{q}_{HX} = \rho c_p v \left( T_{out} - T_{in} \right)$,
where  $\rho$ is the density of air, $c_p$ is the specific heat of air, $v$ is the air speed, and $T_{\rm out}$ and $T_{\rm in}$ are the temperatures of the air entering and leaving the heat exchanger.  Even if $\dot{q}_{HX}$ is increased slightly, the linear change in $v$ ensures that the thermostat is brought closer to the true, mixed temperature of the air in the box, not to the temperature of the water.

To utilize the heat capacity of the water and slow down thermal cycling, one needs instead to reduce the air flow speed around the thermometer so that the sensor is more strongly coupled to the water temperature. The air flow profile across the exchanger is highly inhomogeneous and one can place the thermometer in different air flows simply by repositioning it.  A vane anemometer can be used to measure the air flow speed at any location on the heat exchanger's face.  Changing the air speed at the thermometer in this way does not affect the convection coefficient of the heat exchanger at all. We note that tuning by adjustment of thermometers was not used in our experiments, but this could be done in future work.

Relocation of the thermometer can be explored in the new ETP model by defining a new variable 
    $T_{\rm therm} \equiv (1 -f_{Hm}) T_{a} + f_{Hm} T_{w}$,
where $0 \leq f_{Hm} \leq 1$ is a parameter chosen to represent how close the thermometer is to the well-mixed air temperature versus the water temperature.  When $f_{Hm} = 0$, $T_{\rm therm}=T_{a}$ as before; but when $f_{Hm} = 1$, $T_{\rm therm}=T_{w}$ as if the thermometer is in the water.  This equation is evaluated along with the heat flow equations \eqref{tweqn}-\eqref{t2eqn}, and $T_{\rm \rm therm}$ is used by the thermostat to determine when $T_-$ and $T_+$ are reached.  Because most of the heat injection is through the water heater, $T_{a}$ generally does not overshoot $T_{w}$ or $T_{\rm therm}$.  However, $T_{a}$ typically is much lower than $T_{\rm therm}$ during a cooling cycle as the AC cools the air directly.  

In Fig.~\ref{cycle-durations}(b) the full cycle duration for the new ETP model is plotted against $1-f_{Hm}$, the relative coupling of the thermostat to the true air temperature.  Subsequently, experiments were performed on one of the model houses in which the thermometer was positioned at several different locations above the air-water heat exchanger, and the air flow at those locations was measured with an anemometer.  To relate the air speed to an effective $1-f_{Hm}$, we can use the expressions for $\dot{q}_{HX}$, with $T_{out} \equiv T_{\rm therm}$ and $T_{in} \equiv T_{a}$ to find $f_{\rm eff} = h/\rho c_p$. Because $h$ is not measured, the experimental data are scaled to the curve at a single point. The new ETP model was not calibrated to this particular experimental AC, but the trend is similar. 

\section{HIL Testbed Validation} \label{TestbedValidation}
Prior to running controlled HIL experiments, we validated the experimental and simulation testbeds. Specifically, we sought to demonstrate that the experimental testbed characteristics and behavior were consistent with real systems and the simulation testbed sufficiently replicated behaviors and phenomena seen in the experimental testbed.

\subsection{Experimental Testbed Validation}

We conducted open-loop experiments and validated against the same data used to validate the simulation testbed, specifically, one-second interval residential AC submetering data from 47 homes in Austin, TX~\cite{PSI}. We compared the duty cycles, period of cycles, inrush power to steady state power ratio, and variation in power consumption when ACs are on. Additionally, we compared the correlation between power consumption and both outdoor temperature and voltage. For brevity, here we report our comparisons of the variation in power consumption and the power-temperature correlations; full details are available in~\cite{osti_1804326}.

The validation experiments were performed across several temperature setpoints, with different temperature deadbands, with different heating rates, and under different ambient conditions. No phase-locking, coupling, or coincidence in the on/off cycling of units was observed despite the relatively close proximity of the units to each other, so they acted as independent oscillators, i.e., there were no groups of two or more units that always cycled on and off at the same time.

\begin{figure}
    \centering
    \includegraphics[width = 0.5\textwidth] {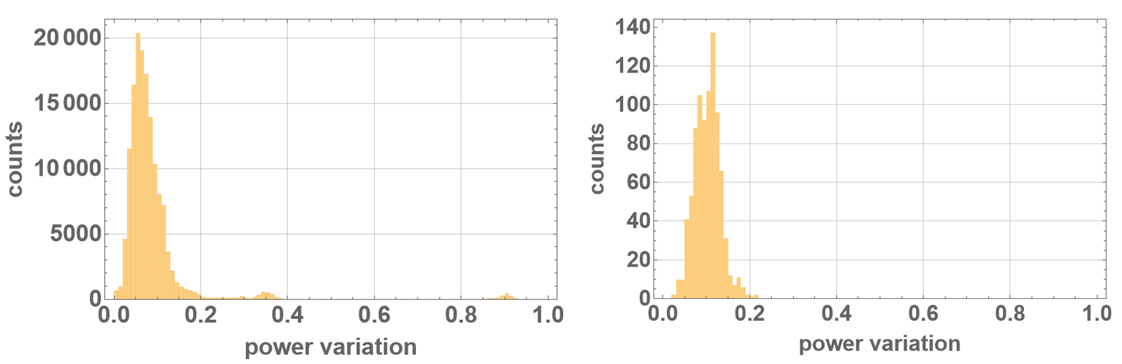}
    \vspace{-.3cm}
    \caption{Fractional power variation when the ACs are on for the experimental ACs (right) and real ACs (left). The distributions approximately match.}
    \label{fig:PowerVariations}
    \vspace{-.5cm}
\end{figure}

For validating the variation in power consumption when ACs are on, we calculate the fractional power variation, defined as the ratio of the difference between minimum and maximum power, and the maximum power. For each on portion of a cycle, we find the minimum and maximum power after discarding the first and last 5~s of data to ensure removal of transients. Histograms of the power variations, for both the experimental ACs and the real ACs, are shown in Fig.~\ref{fig:PowerVariations}. For the experimental ACs, the mean power variation is $10 \pm 3$\% whereas for the real ACs, it is $8.7 \pm 9.3$\%. The larger mean variation of the real ACs was skewed by two ACs with high variation in some cycles. Nonetheless, the power variation distribution of the experimental ACs approximately agrees with that of the real ACs.

\begin{figure}
    \centering
    \includegraphics[width = 0.5\textwidth]{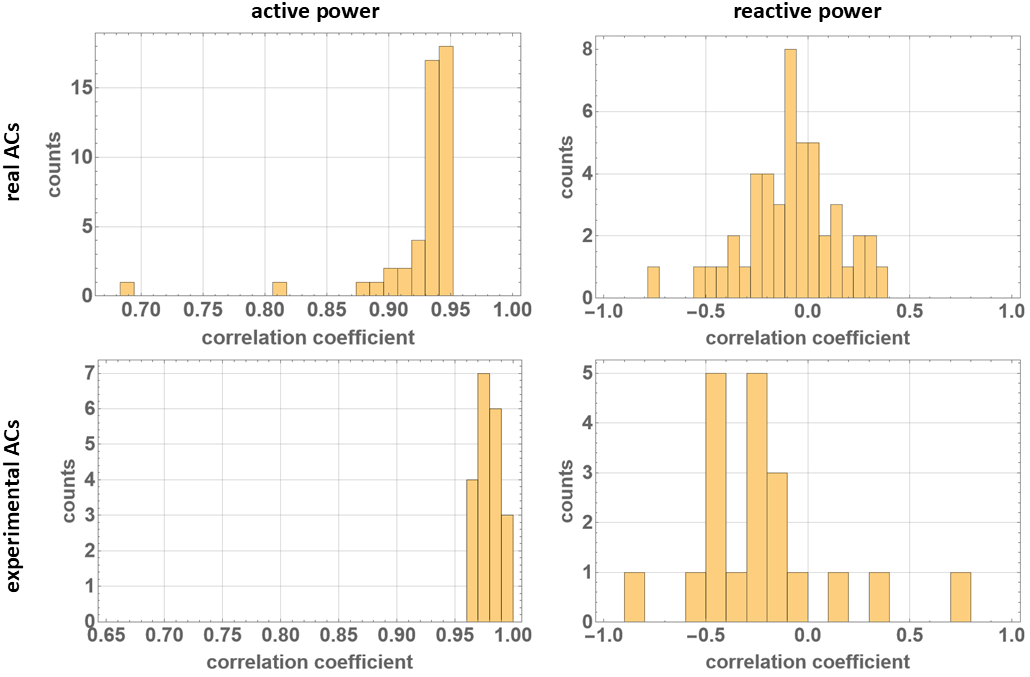}
    \vspace{-.3cm}
    \caption{Active (left) and reactive (right) power to temperature correlation plots for the real ACs (top) and the experimental ACs (bottom). Note the difference in x-axis scale of the histograms for reactive power.}
    \label{fig:PowerTempCorrelation}
    \vspace{-.5cm}
\end{figure}

Fig.~\ref{fig:PowerTempCorrelation} shows histograms of the observed active and reactive power vs.~temperature correlations for both the experimental and real ACs.  For each AC, we extract power and temperature data from when the AC is on, and compute the scalar correlation coefficient  (-1,1) between active power and temperature, and reactive power and temperature. We find that the temperature dependence of power consumption is in fair but not perfect agreement between the experimental and real ACs. This is because the ACs in the testbed are of different models and capacities, and there are no adjustable parameters in the comparison, so perfect agreement could not be expected. The mean fractional change in active power is around 
$1.36\%/^\circ C$ for the experimental ACs and 
$2.12\%/^\circ C$ for the real ACs.

Note that the count difference between the real and experimental data observed in Figs.~\ref{fig:PowerVariations} and~\ref{fig:PowerTempCorrelation} is a function of the limited number of ACs in the experimental dataset (20) versus the real dataset (47) and the shorter time over which the experimental ACs were observed.

\subsection{Simulation Testbed Validation}
Seven open-loop experiments were conducted to verify that any phenomena which are observable in the experimental testbed are also captured and seen in the simulation testbed. The experiments were formulated to induce atypical phenomena, if any, that may not often be observed in practice. For brevity, here we report/discuss comparisons only in terms of power; full details are available in~\cite{osti_1836964}.

The first experiment ran the ACs without external control to explore their open-loop power consumption. The second experiment explored synchronization behavior by controlling all ACs to maximize (or minimize) aggregate power consumption until synchronization, and then releasing control to observe the effects of synchronization on aggregate power variation and the time to desynchronization. The third experiment characterized the frequency response by switching on/off all ACs over a range of different frequencies. The fourth experiment explored on/off switching patterns that could lead to the emergence of multiple synchronous populations of ACs. The fifth experiment explored the impact of AC heterogeneity by repeating experiments 2-4 for homogeneous and heterogeneous populations. The sixth experiment explored the impact of low/high duty cycles by repeating experiments 2-4 for ACs with high and low duty cycles. The seventh experiment explored the impact of communication delays on the degree and/or time to synchronization by repeating experiments 2-4 with deterministic and randomized delays.

\begin{figure}
    \centering
    \includegraphics[width = \columnwidth]{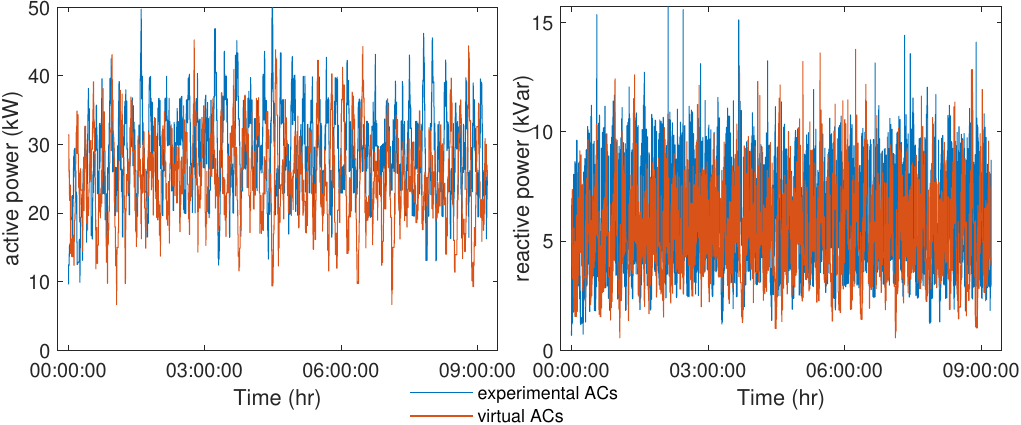}
    \vspace{-.7cm}
    \caption{Active (left) and reactive (right) power comparison between virtual and experimental ACs.} 
    \label{fig:SimExptPowerComparison}
    \vspace{-.3cm}
\end{figure}

Fig.~\ref{fig:SimExptPowerComparison} shows the active and reactive power comparison between 20 virtual ACs and the 20 experimental ACs, obtained in the first experiment. The aggregate power behavior of the virtual and experimental ACs is similar. However, as mentioned in Section~\ref{testbed-behavior}, the effect of delayed temperature dynamics observed in the experimental testbed was significant and not adequately captured in the simulation testbed. Hence, we incorporated these dynamics into the AC simulation models. Instead of using the model developed for understanding the nature of these dynamics in Section~\ref{ETPmodeling} , we incorporated a simpler data-driven model leveraging the real AC data~\cite{PSI}; details can be found in~\cite{temp-delays}.

From the other six experiments, we observed that the behavior of the experimental ACs was diverse even though the model houses were nearly identical.  The heterogeneity of the experimental ACs made it nearly impossible to achieve synchronization, i.e., sustained aggregate power oscillations. Even in the second experiment, when the majority of ACs are forced into the same state by a series of switching commands, releasing the control leads to a rapid de-synchronization of the ACs. This is good news for load control; concerns about AC synchronization may be highly unlikely to materialize in practice. This effect was captured in the simulation testbed by tuning the magnitude and heterogeneity of the AC simulation model parameters. Across all experiments, we found that the simulation testbed behavior was largely in agreement with the experimental testbed behavior.

\section{Hardware-in-the-Loop Experiments} \label{HILexperiments}

We next describe our HIL experiments, with the goals of demonstrating the capabilities and usefulness of our testbed and the ability of aggregations of ACs to provide grid services. 

\subsection{Setup} \label{HILcasestudies}
A single aggregation of both virtual and experimental ACs was coordinated to track power reference signals. To create an aggregation large enough to adequately track signals, we used 523 virtual ACs in addition to the 20 experimental ACs. The parameters of the virtual ACs were generated using GridLAB-D~\cite{GridlabD}, with $\pm20\%$ random variation around nominal values, and each virtual AC consumes 2.6~kW on average when on. The ACs were assigned to buses in taxonomy feeder R5-25.00-1~\cite{PNNL} to explore distribution network impacts.

\begin{table}
    \caption{Simulation Conditions}
    \vspace{-.3cm}
    \begin{center}
\begin{tabular}{ l l l }
\toprule
Condition & Nominal &Extreme \\
\midrule
Reference Signal Type & PJM Reg-D &Square wave\\
Reference Signal Amplitude & 10\% & 20\%, 30\%  \\
Voltage Regulator Setting &Ideal &Non-Ideal\\
Communication Network &Perfect &Imperfect\\
Outdoor Temperature & $90^{\circ}F$ & $100^{\circ}F$ \\
\bottomrule
\end{tabular}
    \label{tab:simulationconditions}
     \end{center}
     \vspace{-.5cm}
\end{table}

Five conditions, shown in Table~\ref{tab:simulationconditions}, were varied to characterize the issues that may arise when AC coordination takes place under nominal and extreme conditions.
``Nominal" conditions use a historical PJM RegD signal~\cite{PJMRegD} as the reference signal, signal amplitudes of 10\% of the aggregation's average power consumption when not controlled (i.e., the baseline power), ideal voltage regulator operation, perfect communication network characteristics (i.e., no communication delays and no packet losses), and an outdoor temperature of $90^{\circ}F$. ``Extreme" conditions use a square wave reference signal (inducing a step-like response), signal amplitudes of 20\% or 30\% of the baseline power, voltage regulator settings that induce over/under-voltages, imperfect communication network characteristics (i.e., delays normally distributed with mean 18~s and standard deviation 3~s, and packet loss across all input commands uniformly distributed between 5 and 10\%), and an outdoor temperature of $100^{\circ}F$. 

It was impractical to run experiments with all possible combinations of conditions above for each controller, hence, a ``Design of Experiments" approach~\cite{Spall2010} informed the experiment selection. Table ~\ref{tab:HILResults} lists the 10 experimental conditions chosen and run for each controller (for a total of 30 experiments). Implementing the extreme outdoor temperature condition in the experimental testbed involved varying the heat gains in the model houses. A heat gain of $200$~W approximated nominal outdoor temperature conditions, while a heat gain of $375$~W was used for the extreme condition.

\begin{table*}
    \caption{HIL Experiment Results for the 3 controllers across the 10 experiments. Nom = nominal, Ext = extreme condition.}
    \vspace{-.3cm}
    \centering
    \begin{tabular}{c|ccccc|c c c|c c c|c c c}
    \toprule
         Case & Signal & Signal & Voltage &Comm. &Out. &\multicolumn{3}{c|}{NRMSE (\%)} &\multicolumn{3}{c|}{PJM Score (0-1)}  &\multicolumn{3}{c}{Transformer overload time (s)}\\
           & Type & Amp. (\%) & Reg. &Net. &Temp.&PI &Markov &PEM &PI &Markov &PEM &PI &Markov &PEM\\  
    \midrule
        1 &Nom & Ext (20) &Nom &Nom &Nom &3.77 & 2.02 &0.87 &0.94 &0.95 &0.97 &235 &282 &222\\
        2 &Ext &Ext (30) &Nom &Nom &Nom &2.76 & 2.26 &1.03 &- &- &- &316 &227 &257\\
        3 &Nom &Ext (30) &Nom &Ext &Nom &13.78 & 11.56 &11.43 &0.89 &0.90 &0.90 &346 &186 &211\\
        4 &Nom &Nom (10) &Nom &Nom &Ext &1.67 & 1.39 &0.57 &0.94 &0.95 &0.96 &582 &502 &526\\
        5 &Nom &Nom (10) &Ext &Ext &Nom &5.55 & 5.33 &5.12 &0.88 &0.87 &0.86 &296 &271 &303\\
        6 &Nom &Ext (30) &Ext &Nom &Ext &4.45 & 2.22 &0.54 &0.95 &0.96 &0.97 &979 &849 &1020\\
        7 &Ext &Nom (10) &Nom &Ext &Ext &2.88 & 2.73 &3.23 &- &- &- &546 &696 &333\\
        8 &Ext &Nom (10) &Ext &Nom &Nom &2.01 & 2.24 &0.94 &- &- &- &315 &301 &177\\
        9 &Nom &Ext (20) &Nom &Nom &Ext &2.82 & 1.67 &0.55 &0.95 &0.96 &0.97 &472 &504 &645\\
        10 &Ext &Ext (30) &Ext &Ext &Ext &5.66 & 5.91 &7.81 &- &- &- &850 &1458 &784\\
    \bottomrule
    \end{tabular}
    \label{tab:HILResults}
    \vspace{-.5cm}
\end{table*}

\subsection{Results and Discussion} \label{HILresults}
Table~\ref{tab:HILResults}
summarizes the experimental results across all cases. We report the Normalized Root Mean Square (NRMSE) error between the aggregate power consumption and the reference signal, normalized by the reference signal mean, for all three controllers across all ten cases. We also report the PJM performance score~\cite{pjm_score} when using the PJM signal as the reference signal, as well as the maximum consecutive overload time duration of transformers in the distribution network. We have not reported over/under-voltage violations because we have found these to be insignificant in our experiments results. 

\begin{figure}
    \centering
    \includegraphics[width=\columnwidth]{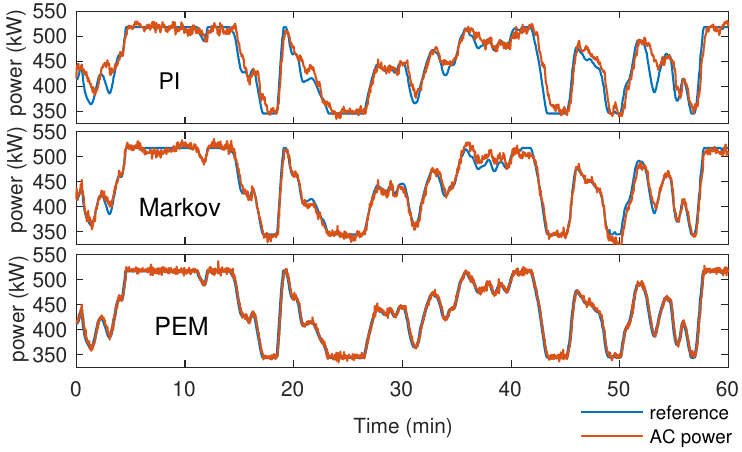}
    \vspace{-.8cm}
    \caption{Comparison of controller tracking performance in Case 1.}
    \label{fig:controller_comparison}
    \vspace{-.5cm}
\end{figure}

From Table~\ref{tab:HILResults} and Fig.~\ref{fig:controller_comparison}, we see that the Markov model-based probabilistic controller and extended PEM controller have better tracking performance than the PI controller in most cases. Also, the PEM controller performs better than the Markov controller in most cases because the Markov controller uses an aggregate model to predict the power consumption of the aggregation and broadcasts probabilistic control inputs, both of which introduce some error, whereas the PEM controller uses direct-from-device turn-on/turn-off requests, which gives the aggregator a precise quantification of power flexibility, and deterministic direct-to-device inputs. The Markov controller's model-based approach provides it with an advantage over the PEM controller in imperfect communication cases 7 and 10.

\begin{figure}
    \centering
    \includegraphics[width=\columnwidth]{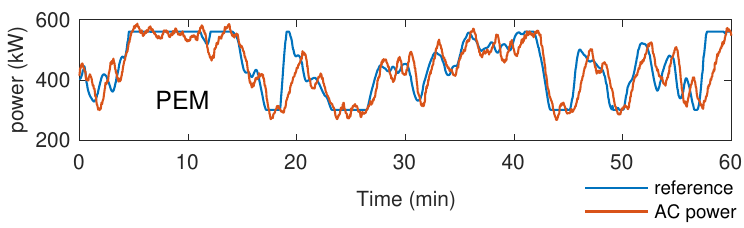}
    \vspace{-.8cm}
    \caption{PEM controller tracking performance in Case 3 (imperfect communication network and larger amplitude reference signal than Case 1).}
    \label{fig:comms_impact}
    \vspace{-.5cm}
\end{figure}

Table~\ref{tab:HILResults} also shows that one of the main factors contributing to significant changes in performance was the communication network. For each controller, cases with extreme communication conditions generally had the largest tracking errors. Fig.~\ref{fig:comms_impact}, which shows PEM controller performance in Case 3 (imperfect communication network and a larger signal amplitude than Case 1), demonstrates the impact of communication delays and packet losses. Comparing the bottom plot of Fig.~\ref{fig:controller_comparison} to Fig.~\ref{fig:comms_impact}, we can see a clear shift of the AC power compared to the reference signal. Comparing cases 1 and 2 where all conditions except signal type and amplitude were held constant, we see that both conditions exert a more benign impact on performance compared to that of the communication condition; this also holds true for outdoor temperature when considering its effect in cases 4, 6, 8, and 9 against other cases with imperfect communication. Note that in the most extreme case (Case 10), the PI controller performs the best, likely because in that case the Markov model is not very accurate and the PEM controller suffers from the imperfect communication network. 

\begin{figure}
    \centering
    \includegraphics[width=\columnwidth]{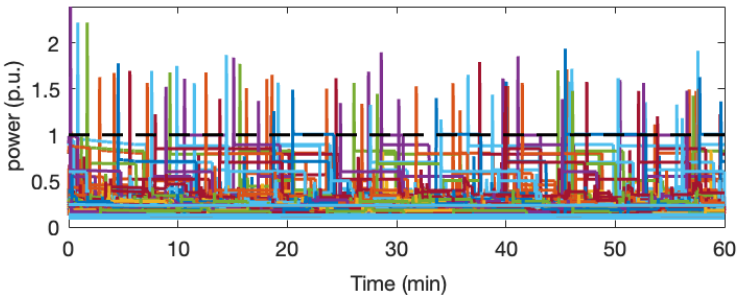}
    \vspace{-.8cm}
    \caption{Transformer power flows in case 1 with the PEM controller. The transformer ratings were 1~p.u.}
    \label{fig:network_impact}
    \vspace{-.5cm}
\end{figure}

Concerning transformer overloading, we see from Table~\ref{tab:HILResults} that there is no clear trend in the maximum transformer overload time duration, leading us to conclude that all the controllers had similar effects on transformer overloading. Fig.~\ref{fig:network_impact} shows the power flowing through the transformers in Case 1 for the PEM controller. Most power flows that exceed transformer ratings (1~p.u.) are instantaneous and/or very small in magnitude. Such overloads do not usually cause actual issues to transformers, as they are not sufficient to cause overheating. In a few cases, the maximum duration is more than a few minutes; however, in these cases,  the overload magnitude was very small (around 1.01~p.u.) and would be unlikely to cause any significant transformer overheating.

\begin{figure}
    \centering
    \includegraphics[width=\columnwidth]{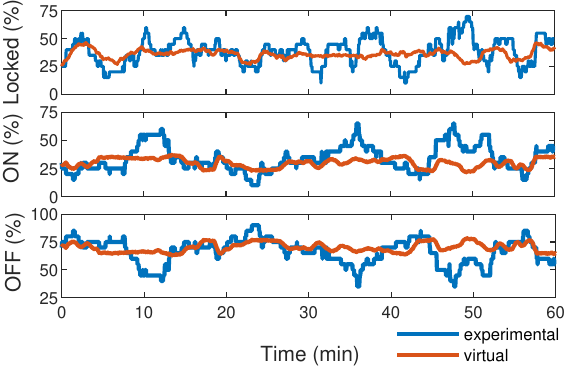}
    \vspace{-.8cm}
    \caption{Percentage of experimental and virtual ACs that are locked (top), on (middle), and off (bottom) for the PEM controller in Case 1.}
    \label{fig:Exp1_TCL_Behavior}
    \vspace{-.5cm}
\end{figure}

We next explore how the experimental ACs fared compared to the virtual ACs. Fig.~\ref{fig:Exp1_TCL_Behavior} compares their lockout and on/off switching behavior in Case 1. The experimental ACs exhibit larger fluctuations due to their much smaller number, but the virtual ACs exhibit fairly similar lockout and on/off switching rates on average, indicating that the experimental ACs were not treated significantly differently than the virtual ACs. Another way to compare experimental and virtual ACs is to calculate the variance of the AC power from the reference signal for select groups of of 20 virtual ACs and compare the variances of those groups to the variance of the experimental ACs. We find that the experimental ACs have a variance that is generally smaller but within the range of variances of the virtual ACs, indicating that the experimental ACs are following the reference signal as well as any group of 20 virtual ACs.  

Overall, from Table~\ref{tab:HILResults}, we observe satisfactory tracking performances across most cases, even the extreme cases. We see the NRMSEs are less than 5\% in most cases and 5-15\% in a small number of cases, mainly under communication network conditions. A PJM performance score of 0.75 is required for market participation, and we see performance scores greater than 0.85 in all cases, even extreme cases, and across all controllers. Also, no oscillations, synchronization, or chaotic behavior was observed in both the extreme open-loop experiments and extreme controlled HIL experiments. These results indicate the feasibility of these load control strategies working in practical settings.

\section{Conclusions} \label{conclusion-section}
We have constructed a HIL experimental testbed with 20 physical model houses enabling us to test load control methods. The model houses provided data on power usage and temperature response that were used to improve high-fidelity AC simulation models. We have shown that the experimental testbed can be tuned and described how we conducted experimental testbed validation against real-world data. We performed HIL experiments to use ACs for frequency regulation leveraging three load control strategies from the literature and demonstrating that ACs are capable of providing fast timescale grid balancing services. Throughout the paper we have highlighted the opportunities and challenges associated with testing AC load control strategies through physical experiments. In particular, experimental testing allowed us to push the system and the controller to the extreme to understand what might happen in practice under worst-case scenarios, which is usually not possible during field testing. 

Future work also includes developing delay-aware controls, because we see from our results that communication delays and packet loss have the most significant impact on performance, and the tested control strategies do not directly address this. We also acknowledge that while it was challenging to construct 20 model houses, the size remains small relative to what would be needed in practice. Hence, further testing on a larger aggregation of experimental devices might be required before such technologies could roll out at scale.

\vspace{-0.07cm}
\appendices
\section{Simple AC model}
\label{appendix}
This appendix describes the simple model of a window AC used to extend the ETP model.  The AC is considered as a Carnot heat pump with extra losses. The power consumption is $\dot{W}_{\rm AC} = \gamma \dot{Q}_c (T_2-T_1)/T_1 + \dot{W}_{\rm fric}$, where $\gamma$ is a loss factor (e.g., internal heat leak), $\dot{W}_{\rm fric}$ is an extra constant loss (e.g., friction), and the other terms are defined in Section~\ref{ETPmodeling}. This model contributes a time-varying active power consumption when the AC is turned on. The compressor in the AC works at a single speed and has only binary on/off states.  When on, the compressor has a fixed rate of displacement, $\dot{V}$, for the working fluid, refrigerant R410A.  The mass flow then depends on the density of the refrigerant at the inlet to the compressor, i.e.,
$\dot{m}=\rho_c(T_1) \;\dot{V},$
where the density $\rho_c$ is a function of the temperature of the cold heat exchanger, from which the heat is being pumped. Although the refrigerant is cold, because it is the material evaporated from the cold heat exchanger, we approximate the vapor as an ideal gas so that   $\rho_c = P_c/R T_1,$ where $R=R_{\rm univ}/M_{\rm mol}$ is the molecular weight-dependent gas constant for this material.  The vapor pressure for the fluid is approximated as $ P_c = k \exp\left( -L/RT_1 \right),$ where $k$ is a constant and $L$ is the latent heat for the refrigerant liquid-vapor transition, which is typically a weak function of temperature, and so taken to be constant in this approximate model.  The value for $L$ was chosen such that $P_c$ best matches the vapor pressure curve for the refrigerant around room temperature, up to an amplitude prefactor. Then, $ \dot{Q}_c = \dot{m} \; L$
and putting together the expressions we obtain    $\dot{Q}_c = A \exp\left( -L/RT_1 \right)/T_1 ,$ where $A$ includes $\dot{V}$, $R$, and $k$.   

\bibliographystyle{IEEEtran}	
\bibliography{main}
\end{document}